\documentclass[twocolumn]{aastex61}
\pdfoutput=1 
\usepackage{amsmath,amstext}
\usepackage[T1]{fontenc}
\usepackage{apjfonts} 
\usepackage[figure,figure*]{hypcap}

\shorttitle{A Dearth of Small Members in the Haumea Family Revealed by the OSSOS Survey}
\shortauthors{Pike et al.}

\begin{document}

\title{A Dearth of Small Members in the Haumea Family Revealed by the OSSOS Survey}

\author[0000-0003-4797-5262]{Rosemary~E. Pike}
\affiliation{Institute of Astronomy and Astrophysics, Academia Sinica; 11F of AS/NTU Astronomy-Mathematics Building, No. 1 Roosevelt Rd., Sec. 4, Taipei 10617, Taiwan}
\author[0000-0002-1788-870X]{Benjamin C. N. Proudfoot}
\affiliation{Brigham Young University, Department of Physics and Astronomy, N283 ESC, Provo, UT 84602, USA}
\author[0000-0003-1080-9770]{Darin Ragozzine}
\affiliation{Brigham Young University, Department of Physics and Astronomy, N283 ESC, Provo, UT 84602, USA}
\author[0000-0003-4143-8589]{Mike Alexandersen}
\affiliation{Institute of Astronomy and Astrophysics, Academia Sinica; 11F of AS/NTU Astronomy-Mathematics Building, No. 1 Roosevelt Rd., Sec. 4, Taipei 10617, Taiwan}
\author[0000-0001-7206-6816]{Steven Maggard}
\affiliation{Brigham Young University, Department of Physics and Astronomy, N283 ESC, Provo, UT 84602, USA}

\author[0000-0003-3257-4490]{Michele T. Bannister}
\affiliation{Astrophysics Research Centre, School of Mathematics and Physics, Queen's University Belfast, Belfast BT7 1NN, UK}
\author[0000-0001-7244-6069]{Ying-Tung~Chen}
\affiliation{Institute of Astronomy and Astrophysics, Academia Sinica; 11F of AS/NTU Astronomy-Mathematics Building, No. 1 Roosevelt Rd., Sec. 4, Taipei 10617, Taiwan}
\author[0000-0002-0283-2260]{Brett J. Gladman}
\affiliation{Department of Physics and Astronomy, University of British Columbia, Vancouver, BC V6T 1Z1, Canada}

\author[0000-0001-7032-5255]{JJ Kavelaars}
\affiliation{NRC-Herzberg Astronomy and Astrophysics, National Research Council of Canada, Victoria, BC, Canada}

\author[0000-0001-8221-8406]{Stephen Gwyn}
\affiliation{NRC-Herzberg Astronomy and Astrophysics, National Research Council of Canada, Victoria, BC, Canada}

\author[0000-0001-8736-236X]{Kathryn Volk}
\affiliation{Lunar and Planetary Laboratory, The University of Arizona, 1629 E University Blvd, Tucson, AZ 85721}

\begin{abstract}

While collisional families are common in the asteroid belt, only one is known in the Kuiper belt, linked to the dwarf planet Haumea. The characterization of Haumea's family helps to constrain its origin and, more generally, the collisional history of the Kuiper belt. However, the size distribution of the Haumea family is difficult to constrain from the known sample, which is affected by discovery biases. Here, we use the Outer Solar System Origins Survey (OSSOS) Ensemble to look for Haumea family members. In this OSSOS XVI study we report the detection of three candidates with small ejection velocities relative to the family formation centre. The largest discovery, 2013 UQ$_{15}$, is conclusively a Haumea family member, with a low ejection velocity and neutral surface colours. Although the OSSOS Ensemble is sensitive to Haumea family members to a limiting absolute magnitude ($H_r$) of 9.5 (inferred diameter of $\sim$90~km), the smallest candidate is significantly larger, $H_r=7.9$. The Haumea family members larger than $\simeq$20~km in diameter must be characterized by a shallow $H$-distribution slope in order to produce only these three large detections. This shallow size distribution suggests that the family formed in a graze-and-merge scenario, not a catastrophic collision.

\end{abstract}

\section{The Haumea Family}
 
The Haumea family was identified by the water-ice surfaces and orbital parameters of its members \citep{brown2007}. 
Haumea family members have a limited range of semi-major axes ($a$), eccentricities ($e$) and inclinations ($i$), unless modified by resonance with Neptune \citep{ragozzine2007}.
In addition to the tightly constrained orbital parameter space, Haumea family members have neutral colours, flat phase curves with high albedo surfaces \citep{rabinowitz2008} and strong water-ice spectral features \citep{brown2007,carry2012}.
A compelling explanation for these surface properties is that the Haumea family members are fragments of Haumea's icy mantle, ejected during a collision1 approximately 3.5 $\pm$ 2 Gyr ago \citep{ragozzine2007,volk2012}. 
The type of collision that created the Haumea family will define the velocity distribution, orbital parameters and size distribution of the family members. 
Proper orbital elements of potential family members are calculated by backwards integration and used for dynamical classification of non-resonant, stable objects, as in asteroid belt family analysis \citep{ragozzine2007}.
The identification of new family members from their orbits and ejection velocities provides a larger sample and can extend the family member identification to objects too small and faint for spectroscopic identification. 
Asteroid families have been found to have size distributions different from the background objects6, so the size distribution of the Haumea family may or may not match the size distribution of the hot classical trans-Neptunian object (TNO) population. 
Haumea family formation models \citep{schlichting2009,leinhardt2010,proudfoot} include a range of relative velocities for the impactors and predict different size distributions for the family, defined using slopes $q$ (diameter distribution) or $\alpha$ ($H$-magnitude distribution). 
Several size distributions including $q = 1.5$ ($\alpha = 0.1$) based on the graze-and-merge formation scenario \citep{leinhardt2010}, $q=3.8$ ($\alpha=0.56$) \citep{fraser2009}, and $q=2.5$ ($\alpha=0.3$) \citep{dohnanyi1969}, have been found to be consistent with the uncertainty for previously known family members \citep{carry2012}. 
These models could previously only be compared to qualitative fits to the apparent size distribution of the bright Haumea family members, such as $q=2$ ($\alpha=0.2$) \citep{lykawka2012}. 
We utilize the detections and survey characterization of the OSSOS Ensemble to constrain the size distribution of the Haumea family.
 
\section{Comparison of Haumea Family Detections and Models}

We searched for Haumea family members in the OSSOS Ensemble, which includes the Outer Solar System Origins Survey \citep{ossos2018}, the Canada-France Ecliptic Plane Survey (CFEPS) \citep{cfeps}, CFEPS High Latitude component (HiLat) \citep{hilat}, and the Alexandersen survey \citep{alexandersen2016}.
The backwards integration using REBOUND \citep{rein2012} identified three Haumea family member candidates with small ejection velocities ($\Delta v$) relative to the family formation center, ($a$=43.10 AU, $e$=0.118, $i$=28.2$^{\circ}$) \citep{ragozzine2007}.  
These include one secure family member and two additional possible family members, listed in Table \ref{targets}.
The secure family member is 2013 UQ$_{15}$, discovered in the OSSOS L block, with a low $\Delta v$ =37 m s$^{-1}$ and neutral surface colours from photometry in the optical and near-infrared filters $g$,  $r$,  $z$ and J;  $g-r$=0.55$\pm$0.02, $r-z$=0.26$\pm$0.16, $r-$J=0.89$\pm$0.11 \citep{pike2017z, colossos} consistent with expectations for a Haumea family member.
Contamination within $\Delta v$ $<$100 m s$^{-1}$ is quite low \citep{ragozzine2007,proudfoot}.
We consider two additional objects with $\Delta v$ $<$160 m s$^{-1}$ as possible family members: 2011 UK$_{412}$ \citep{alexandersen2016} and 2007 RX$_{326}$ \citep{hilat}, however, their smaller sizes and larger $\Delta v$ values increase the likelihood that these may be interlopers and not family members \citep{proudfoot}, and we do not have surface color measurements.

\begin{figure}
\includegraphics[width=.48\textwidth]{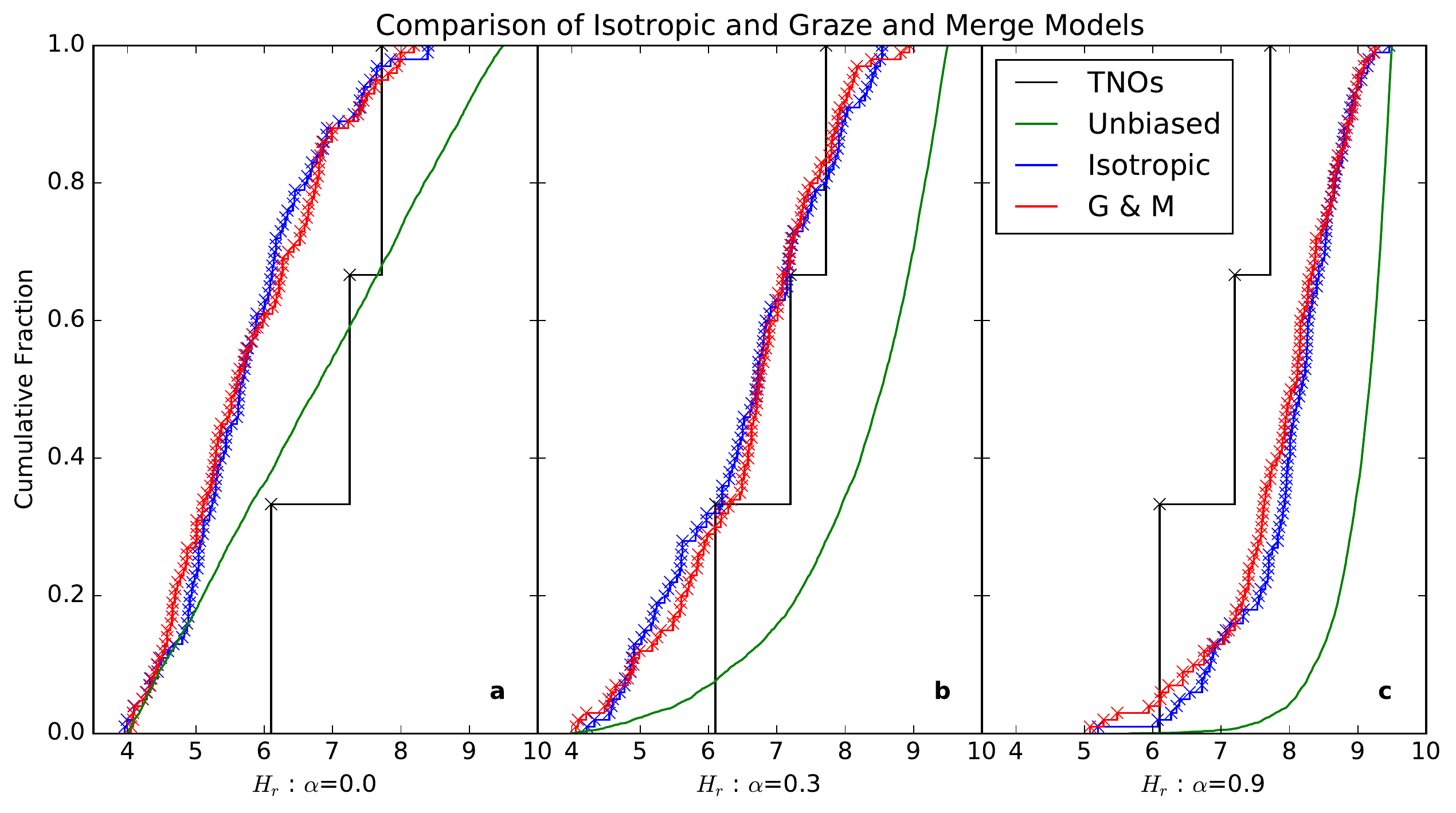}
\caption{Comparison of Isotropic and Graze and Merge models.
The Isotropic and Graze and Merge (G \& M) orbital models$^{11}$ were assigned $H$-magnitudes based on a single slope distribution with $\alpha=0.0$, $\alpha=0.3$, and $\alpha=0.9$ (green).  These models were biased using the survey simulator, and the biased Isotropic (blue) and Graze and Merge (red) do not produce significantly different biased $H$-distributions.  Both provide a good match for the three Haumea family members from the OSSOS Ensemble (black).  We present only the isotropic model results in this work because this model is preferred for the orbital distribution \citep{proudfoot}, and the choice of model does not affect our final results.
\label{isoGM}}
\end{figure}

\begin{table*}
\setlength{\tabcolsep}{2.5pt}
\caption{OSSOS Ensemble and Pan-STARRS1 Survey Potential Haumea Family Members with $\Delta v$ $<$230 m s$^{-1}$}
\label{targets}
\begin{center}
\begin{tabular}{ l l l l l l l}
Survey ID &   MPC ID & Survey & $\Delta v$  (m~s$^{-1}$) & H${_V}$ &H${_r}$ &  Notes\\\hline\hline
 \multicolumn{2}{l}{OSSOS Ensemble} & &  & (mag) & (mag)\\\hline
 o3l77 &  2013 UQ$_{15}$ & OSSOS \citep{ossos2018} & 37 & 6.5 & 6.10 & Very Likely Family Member  \\
 \\
 mal24 &  2011 UK$_{412}$ & Alexandersen \citep{alexandersen2016} & 155 & 7.9 & 7.72 & Possible Family Member \\
 HL7p3 & 2007 RX$_{326}$ & HiLat \citep{hilat} & 158 & 7.5 & 7.25 & Possible  Family Member\\
 \\
o3e29 & 2013 GO$_{137}$ & OSSOS \citep{ossos2018} & 179 & 7.3 & 7.09 & Possible Family Member; beyond $\delta v$ cut \\
 HL7c2 & 2007 FM$_{51}$ & Hilat \citep{hilat} & 190 & 6.7 & 6.59 & Possible Resonant Family Member ($\delta v \simeq 129 \textrm{m~s}^{-1}$); \\
 &&&&&&not currently resonant\\
 
o5d031PD & 2010 VV$_{224}$ & OSSOS \citep{ossos2018} & 225 & 5.9 & 5.80 & Likely Resonant Family Member ($\delta v \simeq 70 \textrm{m~s}^{-1}$); \\
 &&&&&&OSSOS classifies as 16:9 resonant \\\hline
 \multicolumn{2}{l}{Pan-STARRS1} & &  \\\hline
 & 2014 LO$_{28}$ & PS $wgri$ & 14 & 5.2 && Very Likely Family Member \\
& 2014 YB$_{50}$   & PS $wr$  & 106 & 5.8  && Likely Family Member \\
& 2015 FN$_{345}$ & PS $w$ & 125 & 6.0 && Likely Family Member \\
(471954) & 2013 RM$_{98}$ & PS $w$  & 132 & 5.4  && Likely Family Member \\
 &2014 QW$_{441}$ & PS $wgri$ & 137 & 5.0 && Likely Family Member\\ 
&2014 BZ$_{57}$& PS $w$ & 153 & 4.8 && Possible Family Member\\
\\
(499514) &   2010 OO$_{127}$  & PS $gri$ & 70  & 4.6  && Very Likely Resonant Family Member ($\delta v \simeq 50 \textrm{m~s}^{-1}$); \\
 &&&&&&affected by 5:3 resonance \\
&2010 VV$_{224}$& PS $wgri$ & 220 & 5.9 && Likely Resonant Family Member ($\delta v \simeq 70 \textrm{m~s}^{-1}$); \\
 &&&&&&affected by 16:9 resonance\\ 
&2014 XS$_{40}$& PS $wri$ & 226 & 5.3 && Possible Resonant Family Member ($\delta v \simeq 100 \textrm{m~s}^{-1}$); \\
 &&&&&&affected by 5:3 resonance\\
&2015 AJ$_{281}$ & PS $ri$ & -- & 4.9 && Possible Family Member; large uncertainties  \\
\hline
\end{tabular}
\end{center}
\footnotesize{Note.  
Columns include the different object IDs, the survey the TNO was detected in, the $\Delta v$ \citep{proudfoot}, the absolute magnitude in $V$ and $r$ band where reported.  The Pan-STARRS1 Survey $wgri$ observations are non-simultaneous.
The `Notes' column describes whether the TNO is classified as a Family member. 
The $\delta v$ values \citep{ragozzine2007} are ejection velocities corrected based expected orbital modification in resonance.
 Typical uncertainties in $\Delta v$ are $\sim$3\%. \\
}
\end{table*}

The OSSOS Ensemble detections and survey characteristics were used to test for acceptable Haumea family population models (see Methods). We used size distribution models from the literature (listed in Table \ref{Hdist}) and an isotropic model of the population \citep{proudfoot}.
We utilized the isotropic model because it provides the best match to the orbital distribution of the family \citep{proudfoot}. 
Regardless of the orbital distribution model used, the $a$, $e$ and $i$ distribution is tightly constrained. 
We demonstrate in Fig. 1 that the choice of the orbital distribution has a minimal effect on the measured absolute magnitude distribution, and other plausible models of the Haumea family orbital distribution would produce the same result. 
The orbital and $H$-distribution models are compared to the securely classified $\Delta v$ $<$100 m s$^{-1}$  Haumea family member and the three $\Delta v$ $<$160 m s$^{-1}$  likely members. 
The OSSOS Ensemble provided constraints on the faint end of the $H$-distribution, and the Pan-STARRS1 family member candidates provide constraints on the bright end of the distribution (Table \ref{targets}). 
The OSSOS Ensemble can also be used to determine an absolute population number, as the survey detection efficiency is well known.

\begin{figure*}
\includegraphics[width=1.\textwidth]{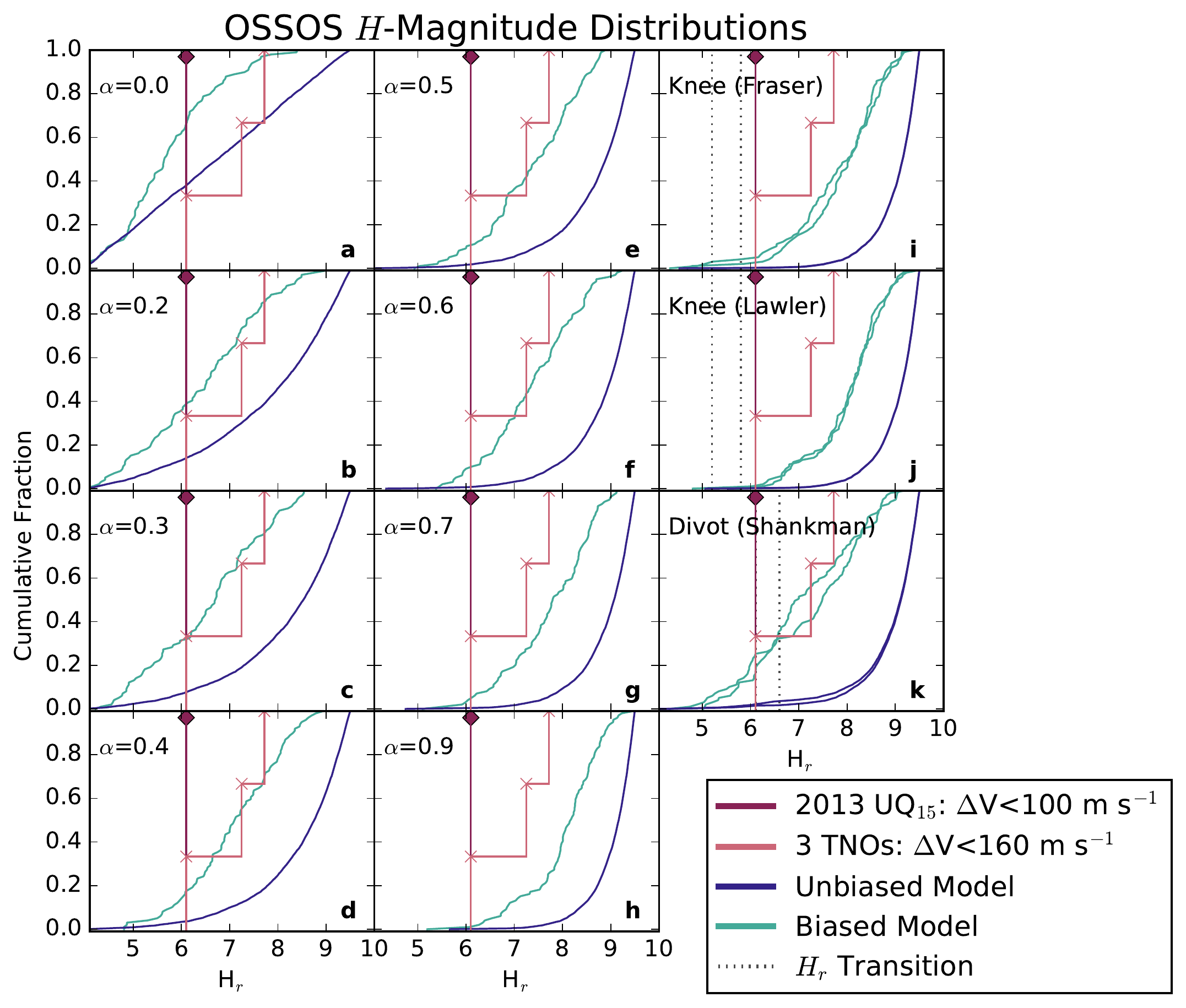}
\caption{Absolute Magnitude Distributions compared to OSSOS Ensemble Detections.
The theoretical size distributions (blue) have either a single slope of $\alpha$, a knee distribution, or a divot distribution (as written in the panel, see Table \ref{Hdist} for details).  The two different transitions in each of the knee and divot distributions (based on different albedo assumptions) have a minimal effect on the biased $H$-distributions.  The survey simulator was used to determine 100 model objects the OSSOS Ensemble surveys would have detected (biased model--turquoise).  The real detected Haumea family member with $\Delta v$ $<$100 m s$^{-1}$ is shown in magenta (`diamond'), and the three likely Haumea family members with $\Delta v$ $<$160 m s$^{-1}$ are shown in light pink (`x').  The $H$-magnitude of 2013 UQ$_{15}$ is brighter than 95\% of the biased model detections (2$\sigma$ rejectable) for a single slope with $\alpha$=0.7--0.9, the Fraser \citep{fraser2014} knee, and the Lawler \citep{lawler2018} knee.  
The three objects with $\Delta v$ $<$160 m s$^{-1}$ also support a shallow slope, rejecting $\alpha\ge0.8$.
Our preferred $H$-distribution is a single slope $\alpha$=0.3.
The large number of 8$<H_r<$9 simulated detections for large $\alpha$ indicates that the Haumea family must have a very shallow slope in this size range.
\label{cumulative}}
\end{figure*}

\begin{table*}
\setlength{\tabcolsep}{2.5pt}
\small
\caption{$H$-distribution Model Parameters and Rejectability}
\small
\label{Hdist}
\begin{center}
\begin{tabular}{ l l  c  c  c  c  c  c  c }
Source & $H$-Distribution &   $\alpha_{\rm bright}$ & $\alpha_{\rm faint}$ & $H_{r-\rm transition}$ & $c$ & OSSOS $\Delta v$ $<$100 m s$^{-1}$ & OSSOS $\Delta v$ $<$160 m s$^{-1}$ & Pan-STARRS1 \\
&Type &&&&&{\bf AD Results} & {\bf AD Results} & {\bf AD Results} \\ \hline
Fraser$^1$ (hot TNOs) & Knee & 0.87 & 0.2 & 5.2 / 5.8 & -- & {\bf 4.9 / 2.6\%}& 8.0 / 19.6\%  & --  \\ 
Lawler$^2$ (scattering TNOs) & Knee & 0.9 & 0.4 & 5.2 / 5.8 & -- & {\bf 1.3 / 0.73\%}& \textbf{3.4 / 2.5}\%  & --  \\ 
Shankman$^3$ (scattering TNOs)&Divot & 0.8 & 0.5 & 6.1 / 6.7 & 5.6 & 25.3 / 17.1\% & 64 / 86\%  & --  \\
  & Single Slope & 0.9 & -- & -- & -- & {\bf 1.2\%} &  \textbf{3.5}\% & {\bf 0.91\% }  \\ 
 & Single Slope & 0.8 & -- & -- & -- & \textbf{2.2}\%&  \textbf{3.2}\%  & \textbf{3.2}\%   \\ 
& Single Slope & 0.7 & -- & -- & -- & \textbf{4.9}\% & 23\%  & 12\%  \\ 
& Single Slope & 0.6 & -- & -- & -- & 10\% & 71\%  & 38\%  \\ 
& Single Slope & 0.5 & -- & -- & -- & 11\% & 94\%  & 54\%  \\ 
& Single Slope & 0.4 & -- & -- & -- & 17\% & 93\%  & 50\%  \\ 
& Single Slope & 0.3 & -- & -- & -- & 32\% & 68\%  & 80\%  \\ 
& Single Slope & 0.2 & -- & -- & -- & 39\% & 59\% & 89\%  \\ 
& Single Slope & 0.1 & -- & -- & -- & 39\% & 53\%  & 53\%  \\ 
& Single Slope & 0.0 & -- & -- & -- & 66\% & 5.5\%  & 43\%  \\ 
\hline
\end{tabular}
\end{center}
\footnotesize{Note.  The Source column indicates where a particular slope was published and which population was studied to measure this slope.  The different $H$-distributions are defined by their slopes $\alpha$ as in equation (\ref{heq}) in the Supplementary Methods.  The broken size distributions utilize two slopes, a transition point $H_{r-\rm transition}$, and can use a contrast $c$ in the case of a divot size distribution \citep{shankman2013}.  Two transitions are included, calculated based on the different reported albedos of the Haumea family. The OSSOS rejectability of each size distribution gives the percentage of biased detections with larger $H$-magnitudes than the OSSOS detections.  The AD statistic was also used to determine rejectability of the different slopes compared to the Pan-STARRS1 detections.  The \textbf{bold} AD results are rejectable at 2$\sigma$ significance.
\\$^1$ \citet{fraser2014}
\\$^2$ \citet{lawler2018}
\\$^3$ \citet{shankman2013}
}
\end{table*}

\section{The Absolute Magnitude Distribution}

\begin{figure}
\includegraphics[width=.5\textwidth]{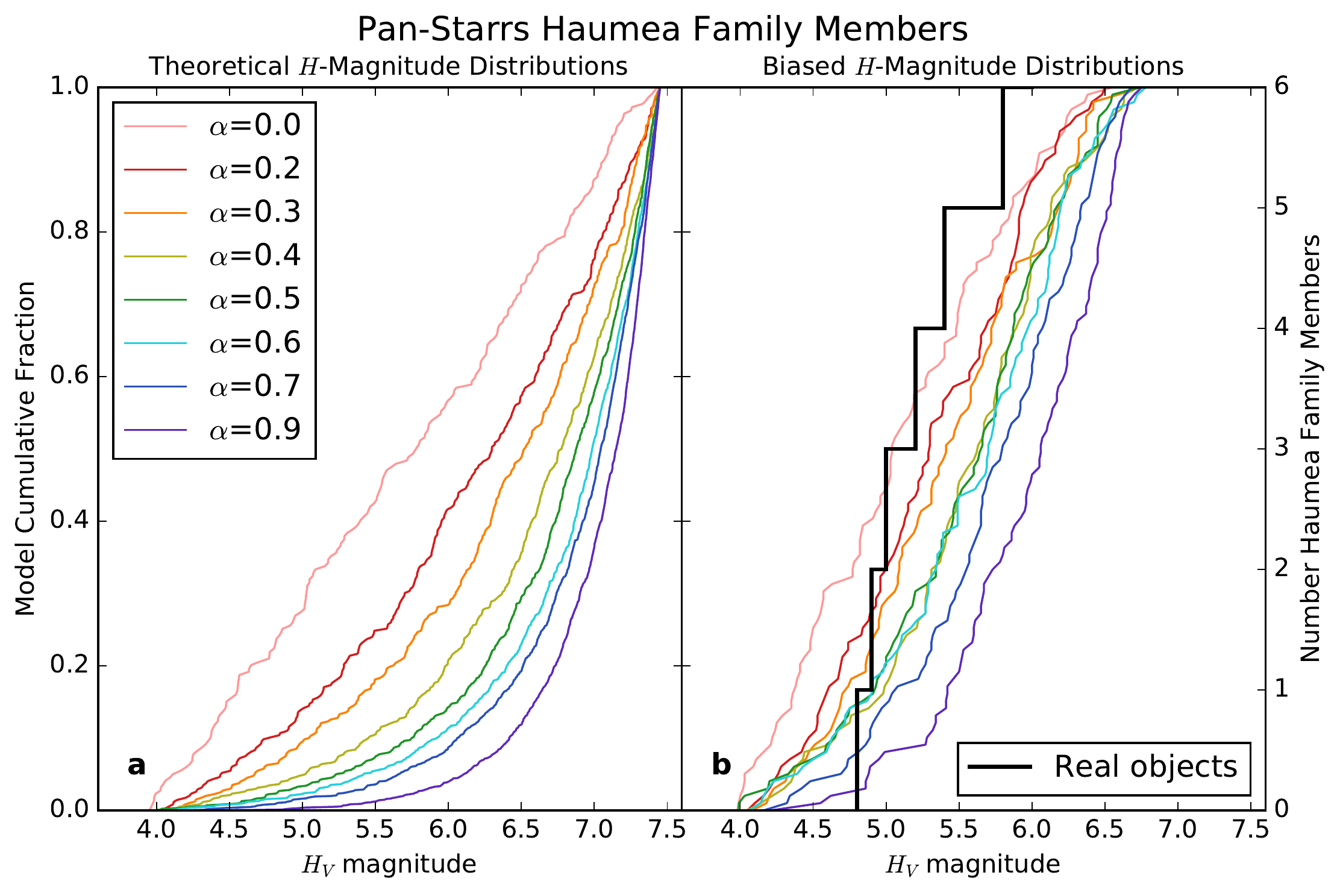}
\caption{Absolute Magnitude Distributions compared to Pan-STARRS1 Detections.
The $H$-magnitude plotted is the Johnsons $V$ band reported in the MPC; $V-r=0.45$.  The scaling (absolute number) is not constrained as these samples do not have a known completeness fraction.
For the theoretical $H$-distribution models, we considered 4$<H_V<$8 as this is the size range of known Haumea family members detectable by the Pan-STARRS1 survey.
{\bf a.} The theoretical size distributions from $\alpha=0.0$ to $\alpha=0.9$.
{\bf b.}  The theoretical size distributions from $\alpha=0.0$ to $\alpha=0.9$ biased using the survey simulator to replicate the effect of the limiting magnitude of the Pan-STARRS1 survey.  These can be directly compared to the cumulative number of Haumea family members (black) observed by the Pan-STARRS1 survey.  The slopes $0.0\le\alpha\le0.7$ provide an acceptable match for the real detections based on the AD statistical test, and shallower slopes are less rejectable.
\label{panstarrs}}
\end{figure}

The preferred literature models for other hot TNO populations do not provide a satisfactory model of the $H$-distribution of the Haumea family.
Both the OSSOS Ensemble survey detection and the Pan-STARRS1 detections, which span different $H$-magnitude ranges, favor a shallow single slope; we find no evidence for an absolute magnitude (or size) distribution transition.
The literature models show a poor visual match to the real detections in Figures \ref{cumulative} and \ref{panstarrs}, which is quantified using the Anderson-Darling (AD) \citep{andersondarling54} statistical test in Table \ref{Hdist}.
A variety of single slope size distributions provide a good match; they are not rejected by the AD statistic for 1-3 OSSOS detections and also provide an acceptable match to the Pan-STARRS1 TNOs.
A change of slope (seen in other TNO populations) is not necessary to fit both the bright and faint end of the size distribution in the range of magnitudes detected by these surveys.
The size distribution of the Haumea family members is well represented by a single slope $0.1\leq\alpha\leq0.4$, although $0.0\leq\alpha\leq0.6$ is not statistically rejectable.
Steeper slopes are entirely rejectable by both the OSSOS Ensemble and the Pan-STARRS1 detections. 
Our preferred slope is $\alpha =  0.3$, which provides a convincing match to both the OSSOS Ensemble and the Pan-STARRS1 survey.

The OSSOS Ensemble surveys demonstrate clearly and conclusively that there is a lack of small members of the Haumea family population.
The single OSSOS Ensemble detection at $H_r=6.1$ with $\Delta v$ $<$100 m s$^{-1}$ is three magnitudes brighter than the detection limits for these surveys.
Although the wider cut of $\Delta v$ $<$160 m s$^{-1}$ includes two additional possible family members, the three detections are still best represented by a shallow $H$-distribution.
Steep slopes of $\alpha\ge0.7$ are ruled out at 2$\sigma$ significance for the Haumea family, because the survey simulator determines that the $H_r$=6.1 detection should have been accompanied by $\sim10$ fainter Haumea family members if their size distribution is characterized by a steep slope.
With a slope of $\alpha$=0.3 , the Haumea family size distribution is more similar to the post-transition slope of the hot population (0.2--0.4 \citep{fraser2014,lawler2018}) than the bright-end distribution.

The Pan-STARRS1 detections also favor a shallow absolute magnitude distribution.
The six Haumea family members detected by the Pan-STARRS1 survey are all brighter than the completeness limit ($m_V=22.5$), and thus provide the best sample currently available for measuring the slope of the bright-end $H$-distribution of the Haumea family.
Size distributions with $\alpha\geq0.8$ are statistically rejectable based on these detections.
Although all slopes $0.0\le\alpha\leq0.7$ were statistically acceptable, $0.1\leq\alpha\leq0.5$ provides a significantly better match to the size distribution of the real detections.
The OSSOS Ensemble's statistically rigorous constraint for the bright Haumea family members is consistent with $\alpha$=0.3 \citep{proudfoot} and $\alpha$=0.44 \citep{vilenius2018}, both of which were based on larger samples with less quantifiable selection effects.
This shallower slope for the bright-end distribution is consistent with the OSSOS detections, and implies that there is no transition in the absolute magnitude distribution in the size ranges that are probed by the Pan-STARRS1 survey and the OSSOS Ensemble, 4$<$$H_r$$<$ 8.
As a result, we determine that a single $H$-distribution with a shallow slope is sufficient to describe the Haumea family $H$-distribution in this $H$-magnitude range.

\newpage
\section{Population Estimate and Ejected Mass Estimation}

Based on the single slope model of the size distribution with a slope $0.0\leq\alpha\leq0.6$, we can estimate the size and mass of the Haumea family.
We combine the isotropic orbital distribution and single slope size distributions with the OSSOS Survey data to determine a population estimate (see Methods).
Because the characterization for the OSSOS Ensemble surveys is significantly more precise than the Pan-STARRS1 survey description, we used the OSSOS characterizations and the one or three OSSOS detections to determine the population estimate for each $\Delta v$ range.
Using the Pan-STARRS1 detections to calculate a population estimate would require a much more precise field locations, detection efficiencies, and observation time definitions than is publicly available.
We calculate population estimates for the stable portion of the Haumea family, which has not been modified by resonant interaction with Neptune and meets the $\Delta v$ $<$100 and 160 m s$^{-1}$ criteria.
A single slope size distribution with $0.0<\alpha<0.6$ is not rejectable by either sample, although we favor shallower slopes as these predict fewer faint detections, consistent with the lack of small objects in the OSSOS Ensemble.
Based on the isotropic model of the spatial distribution of the Haumea family \citep{proudfoot}, the single slope size distribution, the OSSOS Ensemble detections, and the careful characterization of the OSSOS Ensemble, we calculate the range of population estimates that are reasonable for the Haumea family, presented in Table \ref{popEst}.
For the preferred slope of $\alpha$=0.3, we find that the stable Haumea family population with $\Delta v$ $<$160 m s$^{-1}$ and $H_r>9.5$  contains 450$_{-390}^{+720}$ objects with $2\sigma$ confidence.

\begin{table*}
\setlength{\tabcolsep}{2.5pt}
\caption{Population Estimates and Implied Mass of Ejected Fragments}
\label{popEst}
\begin{center}
\begin{tabular}{   c | c  c  c c | c c  c }
 &  & \textbf{Population} &  \textbf{Estimate}  &  & Implied Ejected Mass & [\% of Haumea] \\
Slope  & $\Delta v$ $<$100 m s$^{-1}$ & $\Delta v$ $<$100 m s$^{-1}$ & $\Delta v$ $<$160 m s$^{-1}$ &  $\Delta v$ $<$160 m s$^{-1}$&  $\Delta v$ $<$100 m s$^{-1}$ & $\Delta v$ $<$160 m s$^{-1}$   \\
$\alpha$ & $3.5<H_r<9.5$ & $3.5<H_r<6$ &  $3.5<H_r<9.5$ & $3.5<H_r<6$ & $3.5<H_r<9.5$ & $3.5<H_r<9.5$ \\ \hline
0.6 & 284$^{+1134}_{-293}$  &   3$^{+10}_{-3}$  &  1171$^{+1964}_{-886}$ & 10$_{-7}^{+17}$ & 0.65\% & 1.4\% \\ 
0.5 & 281$^{+11189}_{-273}$   &    6$^{+19}_{-6}$ & 930$^{+1427}_{-710}$  & 16$_{-14}^{+31}$ &  0.84\% & 1.8\%\\ 
0.4 & 200${^{+801}_{-186}}$   &  8$^{+32}_{-7}$  & 671$^{+1007}_{-518}$  & 27$_{-21}^{+40}$ & 1.0\% &  2.5\% \\ 
0.3 & 116${^{+491}_{-107}}$   &   11${^{+44}_{-10}}$  &  436$^{+762}_{-321}$ & 39$_{-34}^{+63}$ & 1.2\% & 3.2\% \\ 
0.2 & 81${^{+341}_{-72}}$  &   16${^{+68}_{-14}}$  &  323$^{+550}_{-247}$ & 64$_{-55}^{+92}$ & 1.5\% & 4.9\% \\ 
0.1 & 53$^{+250}_{-47}$  &   23$^{+111}_{-21}$  & 235$^{+373}_{-188}$  & 104$_{-87}^{+128}$ & 1.9\% & 6.9\%  \\ 
0.0 & 37$^{+209}_{-35}$  &   15$^{+87}_{-14}$  &  169$^{+278}_{-133}$  & $70_{-15}^{+46}$ & 2.3\% & 8.9\%  \\ 
\hline
\end{tabular}
\end{center}
\footnotesize{
The population estimates were calculated for the non-rejectable $H$-distribution slopes and assuming 1 or 3 detections in the OSSOS Ensemble.
The 2$\sigma$ uncertainty on the population estimate is calculated by running the survey simulator 2,000 separate times until 1 (for $\Delta v$ $<$100 m s$^{-1}$) or 3 (for $\Delta v$ $<$160 m s$^{-1}$) objects are detected.  The median value is quoted as the population estimate, and the central 95\% of the values are bounded by the 2$\sigma$ uncertainties.  The $H_r<6$ is calculated by scaling the fainter population estimate using the input $H$-distribution model.  Assuming an albedo of 0.85--0.48, $H_r=3.5$ corresponds to a diameter of 288--383~km, $H_r=6$ corresponds to a diameter of 91--121~km, and $H_r=9.5$ is a diameter of 18--24~km.  
There are 22 candidate family members with 3.95$<H_V<6.45$ for $\Delta v$  $<$160 m s$^{-1}$ and 7 candidate family members for $\Delta v$ $<$100 m s$^{-1}$ \citep{proudfoot}. All of the slopes shown here produce a number of intrinsic objects consistent with the known Haumea family members.
}
\end{table*}
 
We combine the population estimate of the Haumea family with the size distribution model to estimate the mass of the family for both $\Delta v$  assumptions.
We assume that the family members have the density of water ice, $9.34\times10^{11}$~kg~km$^{-3}$.
The mass of the family model is calculated by integrating the differential size distribution and the mass equation \citep{petitSSBN}.
The normalizing constant was determined from the population estimates.
We increase the total mass of the family to account for loss due to instability or resonance occupation, because 30\%$\pm$10\% of the original family is lost or currently resonant \citep{volk2012,lykawka2012}, and these are not included in our family model. 
Some of these are lost into the inner solar system, and a rough estimate is that $\sim$0.1\% of Earth's oceans could have originated from the Haumea family collision.
As it is just outside the modeled $H$ range, a mass for 2002 TX$_{300}$ was also added, assuming a radius of 161~km \citep{vilenius2018}.
We determined the implied ejected mass as a fraction of Haumea's mass ($4.01\times10^{21}$ kg).
For our preferred size distribution, $\alpha=0.3$ or $q=2.5$, with $\Delta v$ $<$160 m s$^{-1}$, the ejected mass is 3\% of the mass of Haumea. 
See Table \ref{popEst} for the mass of the Haumea family for different size distributions. 
Each individual estimate has significant statistical and systematic uncertainties, but for our preferred size distributions, the total mass in the family is a few percent of Haumea's current mass. 
A previous estimate of the total mass in known Haumea family members is $\sim$2.6\% \citep{vilenius2018}, consistent with these results; our shallow size distribution slope implies that most of the mass is in the known large objects. 

\section{Discussion}

The Pan-STARRS1 survey is sensitive to $H_V < 6.5$  and OSSOS extends this sensitivity to $H_V < 9.5$. With the assumed values of albedo $\rho$=0.85--0.48, the TNOs would have diameters of 91--121 km ($H_V = 6.5$) and 18--24 km ($H_V = 9.5$), probing to $\sim$3 times smaller for this population than has been possible for other hot classical TNOs. 
If the Haumea family members had the same size distribution as the rest of the hot TNO populations, the Haumea family would be ideal for probing the size distribution transition. 
However, neither the knee nor the divot size distribution models appropriate for the hot TNO populations provide a good match for the observed distribution of Haumea family members. 
The knee/divot is speculated to represent the transition point between the large-object primordial slope and the collisional remnants.
The similarity of the Haumea family slope to the post-transition slope is suggestive, but the uncertainty in these measured slopes and different transition models can only be resolved with additional discovery surveys (deep surveys to probe the post-transition slope and wide surveys to discover more Haumea family members).
Another possible complication is whether collisional processing affects albedo; if collisions reveal brighter sub-surface material, this will have dramatic effects on the assumed size distribution of the small hot TNO populations, which are typically assumed to have very low albedos.

This shallow size distribution slope creates a challenge for future work characterizing the Haumea family size distribution.
For shallow $H$-distributions, a wider area at a fainter limiting magnitude is more efficient for discovery.
The Large Synoptic Survey Telescope (LSST) \citep{ivezic2008} will have a survey depth of $m_r\sim$24.5, somewhat shallower than OSSOS and deeper than the Pan-STARRS1 Solar System survey, but will cover a larger region of the sky. 
One implication of our results is that the discovery likelihood for Haumea family members with LSST is smaller than would have been previously expected.
Based on the OSSOS population estimates for $\alpha$=0.3, if LSST detects all objects with $\Delta v$ $<$160 m s$^{-1}$ and $H_r\le$7, the survey will find approximately 80 family members (including known objects).
Because of the shallow size distribution, the large-area surveys will provide the best opportunity to better constrain the size distribution in the future.

The shallow slope of the $H$-distribution will have important implications for future modeling of the Haumea family formation and pre-collision state, which can be better explored now that the sample of detected Haumea family members has increased as a result of recent survey and classification efforts.
We find that the shallow slopes $0.2\le\alpha\le0.4$ all provide an excellent match to the Haumea family detections.
With our preferred shallow slope of $\alpha$=0.3, the Haumea population contains 450$_{-390}^{+720}$ members (95\% confidence) with $H_r<9.5$, and a total mass of 3\% of the mass of Haumea.
A recent simplified detectability analysis \citep{proudfoot} on a larger sample of the Haumea family found a consistent result to this work; the family is characterized by a shallow $H$-distribution slope with an upper limit of $\alpha\sim0.4$.
A surface classification analysis \citep{vilenius2018} derives somewhat different ($\alpha=0.44^{+0.1}_{-0.08}$), but not statistically rejectable, distribution than this work, and results in a similar total mass of the family.
It is a challenge for family formation models to match all of the constraints based on observations of the Haumea family, including the orbital distribution (which appears isotropic), absolute magnitude distribution (characterized by shallow slopes), and ejected mass (a few percent of Haumea's mass).
A variety of hypotheses have been proposed for formation of the Haumea family.
Some of these formation models include the catastrophic disruption of a small body \citep{ortiz2012,schlichting2009,campobagatin2016} which, consistent with our results, would have had a mass of at least a few percent of Haumea's mass. 
Based on modern collision simulations, the disruption of an object with 3\% of Haumea's mass would have a typical ejection velocity of $\sim$230 m s$^{-1}$ with a wide dispersion, much larger than the observed velocity distribution of the family \citep{leinhardt2012, proudfoot}.
Thus our measured mass is strongly inconsistent with hypotheses which include the disruption of a small body. 
Furthermore, the shallow size-distribution we detect is not a good match for such collisions, which typically have statistically rejectable slopes of $\alpha$=0.7-0.9 \citep{leinhardt2012}.
The number-size-velocity distribution we detect is a better match to family formation hypotheses that invoke rotational fission \citep{leinhardt2010, ortiz2012}, but these are inconsistent with the distribution of observed proper orbital elements of the entire known family; see Proudfoot \& Ragozzine \citep{proudfoot} for additional discussion. 
Future work on the formation of the Haumea family will require additional models and simulations to determine what type of event can reproduce the shallow size distribution, population size, and the near-isotropic ejection distribution.


\section{Methods}

\subsection{$H$-Distribution Analysis}

The Haumea family population models were tested using the OSSOS Ensemble detections and survey characteristics.
For a detailed description of the use of the Survey Simulator, see Lawler et. al \citep{lawlerFASS}.
The small number of detected Haumea family members mean that few constraints can be provided for the orbital distribution; however, the lack of detections provides important insight into the family's size distribution.
The OSSOS Ensemble is characterized (the detection biases are well understood and modeled) and thus can be used to determine an absolute population number, based on the careful record of survey pointings, sensitivity, and detections.

We tested the OSSOS Ensemble detections against two models of the orbital distributions produced in different collision scenarios: an isotopic distribution, and a `graze and merge' distribution \citep{proudfoot}. 
The isotropic distribution produced by a catastrophic collision best resembles the orbital distribution of the known family members.
The graze and merge scenario produces a more complex orbital distribution, with dependencies between $a$, $e$, and $i$ not seen in the known family members \citep{proudfoot}.
Particles which were not long term stable were removed from the models based on similar models which have determined that 60-80\% of the family is long-term stable \citep{lykawka2012,volk2012}.
While there are some differences between these model distributions, both models have the very similar tight limits on $a$, $e$, and $i$, and those orbital parameter differences are not significant enough for the OSSOS Ensemble detections to differentiate between them.
Figure \ref{isoGM} shows the $H$-distributions that result from three slope($\alpha$) values, the extreme $\alpha=0.0, 0.9$ and the preferred $\alpha$=0.3.
The isotropic and graze and merge models produce nearly identical biased $H$-distributions.
Because the isotropic model is strongly preferred based on all known Haumea family members \citep{proudfoot}, we present the isotropic model results in this work.

The different TNO size distributions and orbital distributions were tested using the OSSOS survey simulator, which uses the survey pointings and sensitivities to determine the detectability of model objects \citep{lawlerFASS}.
We randomly assigned $H$-magnitudes from the different size distributions to model objects from the orbital distribution model.
The survey simulator then determined which model objects were detectable by the surveys; this reveals the sensitivity of the surveys to objects with both different orbital parameters and different $H$-magnitudes (Figure \ref{cumulative}).
The theoretical size distributions (single slope, knee, and divot) with a variety of parameters were assigned to the isotropic Haumea family model and then input into the survey simulator to determine detectability.
The survey simulator was run until it produced 100 detections for each input model.
In this analysis, the real detections can be directly compared to the `detections' as `observed' by the survey simulator because they have the same detection biases.
The biased distributions show the $H$-magnitude distribution that would be measured by the OSSOS Ensemble surveys.
This was compared to both the single detection with $\Delta v$ $<$100 m s$^{-1}$ and the three detections with $\Delta v$ $<$160 m s$^{-1}$ to provide the results for both confidence levels.

A variety of absolute magnitude distributions from the literature were explored for the family members.
Absolute magnitude distributions are used as a proxy for size-distributions, and if the albedo is assumed to be constant they can be scaled directly into diameter distributions.
Typical absolute magnitude and size distributions use a power law with increasing numbers of objects at smaller sizes.
In differential form, the number of objects, $N$, per absolute $H$-magnitude is defined in terms of the slope $\alpha$:
\begin{equation}
\label{h_equation}
  dN/dH \propto 10^{\alpha H} .
\end{equation}
The detection of between one and three Haumea family members in OSSOS restricts the power of the survey to constrain the shape of the $H$-distribution.
However, the characterized survey blocks can test whether the detected objects (and non-detections) are consistent with a proposed $H$-distribution, shown in Figure \ref{cumulative}.
We tested several published $H$-distributions which provide compelling matches to dynamically hot populations in the Kuiper belt; their parameters are reported in Table \ref{Hdist}.
They include a single slope $H$-distribution, knee distributions \citep{fraser2014,lawler2018}, and a divot distribution \citep{shankman2013,lawler2018}.
The distributions are defined using the slope $\alpha$ or $\alpha_{\rm bright}$ and $\alpha_{\rm faint}$ for a joined distribution at a location $H_{\rm transition}$.
For the non-continuous divot distribution, a contrast $c$ is also used \citep{shankman2013}.
The non-continuous divot distribution is included for completeness, however, there is no reason to expect this size distribution shape for a collisional family.
We also tested a single slope size distribution covering the range of slopes predicted in collisional models.

Haumea family members have a significantly higher albedo than typical TNOs, so if the transition magnitude occurs at the same absolute size, this must be shifted in $H$-magnitude for Haumea family members.
Dynamically excited TNOs have been found to have median albedos of $\rho$=0.085$^{+0.084}_{-0.045}$ based on Herschel observations of their flux \citep{vilenius2014}.
A range of albedo measurements have been found for Haumea and its family.
 Recent work has determined the typical albedo for Haumea family members to be $\rho$=0.48 \citep{vilenius2018}, and an albedo for Haumea of $\rho$=0.51 was determined through stellar occultation \citep{ortiz2017} and its satellites were found to have albedos $>0.5$ \citep{muller2018}.
Previous analysis had found $\rho$=0.80 for Haumea \citep{fornaiser2013} and $\rho$=0.88$^{+0.15}_{-0.06}$ \citep{elliot2010,lellouch2013,vilenius2018} for family member 2002 TX$_{300}$.
We test the range of proposed albedo values for Haumea and its family in order to determine whether any of these values produce an absolute magnitude distribution that provides a compelling match to our detections.
The difference in albedo of Haumea family members compared to typical TNOs has a significant effect on the absolute magnitudes of the TNOs.
 The change in magnitude for a given change in albedo is given by:
\begin{equation}
  H_{\rho_1} - H _{\rho_2} = -2.5 \times  \log (\rho_1 / \rho_2) .
  \label{heq}
\end{equation}
To determine the transition $H$-magnitude, we use the median albedo for the dynamically excited TNOs, $\rho_1$=0.085, and an albedo of $\rho_2$=0.85--0.48 for the Haumea family.
A typical Haumea family member of a given size will therefore be approximately 2.5--1.9 magnitudes brighter than a typical hot classical TNO of the same size.
In order to compare the Haumea family objects' $H$-distribution to literature models, the transition magnitude was shifted based on the albedo of the Haumea family (-2.5 and -1.9 magnitudes) and $H_g$ was converted to $H_r$ based on solar colors ($g-r$ = 0.45$\pm$0.02) \citep{holmberg2006}.

In Figure \ref{cumulative}, it is clear that the single OSSOS detection at $H_r$=6.1 is extremely unlikely for the majority of absolute magnitude distributions.
For a single slope with $\alpha\geq$0.7 and the knee distributions \citep{fraser2014,lawler2018}, the detection is brighter than $>95$\% of the biased detections, rejectable at 2$\sigma$ significance.
These results do not change significantly between the higher and lower albedo $H$-transitions.
If we consider all three detections with $\Delta v$ $<$160 m s$^{-1}$, using the Anderson-Darling (AD) statistical test \citep{andersondarling54}, these detections also reject a steep absolute magnitude distribution slope.
These $H$-distributions all imply a large number of small Haumea family members should have been detected by the OSSOS Ensemble survey in addition to the bright object detections.
Based on the OSSOS Ensemble detections, we expect a shallow slope for the size distribution beyond $H_r\sim$6, and prefer a very shallow slope of $\alpha\leq$0.3.
A different knee than those proposed for other TNOs may produce slightly better results, but the results for the Fraser \citep{fraser2014} knee, which uses a very shallow post-transition slope, are not a compelling match.
From the OSSOS results alone, a shallow slope is required beyond $H_r\sim6$, however we cannot constrain whether this slope is the same for the bright end of the $H$-distribution.

To determine a representative bright-end $H$-distribution for the Haumea family, we consider the wider but shallower Pan-STARRS1 `The Solar System Survey' survey, which is the $w$ band portion of the Pan-STARRS survey efforts, focused on finding Solar System objects \citep{denneau2013,magnier2013}.
We consider the bright-end of the $H$-distribution to begin at $H_V\sim4$, as this is the absolute magnitude of the largest Haumea fragments beyond Haumea and 2002 TX$_{300}$.
We identified Haumea family members detectable by the Pan-STARRS1 survey as those in the family list \citep{proudfoot} that had $w$-band Pan-STARRS1 astrometry reported at the Minor Planet Center (MPC, \textit{https://www.minorplanetcenter.net/}), as of June 30, 2018.
These objects are: 2014 LO$_{28}$, 2014 YB$_{50}$, 2015 FN$_{345}$, 2013 RM$_{98}$, 2014 BZ$_{57}$, and 2014 QW$_{441}$, see Table \ref{targets}.
All are within the brightness limit to which Pan-STARRS is reported to be complete ($m_w$=22.5) \citep{lin2016}, or $m_r\sim22.0$.
We exclude 2010 OO$_{127}$, because of its possible resonance modification and lack of observations in $w$ band.
Table \ref{targets} also lists three more possible Pan-STARRS1 Haumea family members, which we exclude because they have $\Delta v$ $>$160 m s$^{-1}$.
The moving object search criteria for the Pan-STARRS1 survey was sensitive to rates of motion from 0.15--15 arcseconds hr$^{-1}$, which includes sensitivity to the Haumea family members.
Because of the tight $a$, $e$, and $i$ constraints on the Haumea population and the lack of longitudinal biases for the non-resonant family members, the only significant bias in the Pan-STARRS1 survey for our target population is the magnitude limit.

To quantify the effect of the completeness limit, we used the OSSOS Survey Simulator to test a very simple model of the Pan-STARRS1 `The Solar System Survey'.
No pointing and depth per observing region is provided by the Pan-STARRS1 survey at this time, however the Solar System Survey covers the majority of the ecliptic plane to $\pm$30$^{\circ}$ to a limiting magnitude of $m_w$=22.5 \citep{lin2016}.
Based on reported Pan-STARRS1 survey limiting magnitude in $w$, we use a limiting magnitude of $m_r\sim22.05$ for Haumea family members for all survey regions.
To constrain the shape of the $H$-distribution (not the absolute scaling), the limiting magnitude is very important, but the precise survey area is not critical, under the reasonable assumption that there is no size-inclination dependence of the objects.
In our orbital model, the objects have no dependence between $H$-magnitude and right ascension-declination position, so the specific location of the simulated block locations does not affect conclusions about the size distribution.
We define a series of large observing blocks across the full right ascension range spanning -30$^{\circ}$ to +30$^{\circ}$ declination.
Each block is assumed to be observed at opposition, where the rate of motion of these objects is sufficiently large to be easily detectable within the Pan-STARRS1 survey rate cuts.
This simple model is sufficient to determine the effect of the magnitude limit on the apparent $H$-distribution of Haumea family members in the Pan-STARRS1 survey.

The survey simulator was run using the isotropic model of the Haumea family and the simple Pan-STARRS1 survey blocks to create simulated detections (Figure \ref{panstarrs}).
Once the size distributions have been appropriately biased based on the completeness, the shallower slopes provide a significantly better match to the real detections.
The AD test was then used to compare the real Haumea family members to the simulated detections and rules out single slope size distributions with $\alpha\geq0.8$.
Similar to the OSSOS Ensemble surveys, the best results are $0.2\leq\alpha\leq0.3$.

\subsection{Population Estimate}

To determine a population estimate, the isotropic orbital model and $H$-distribution are input into the survey simulator, and the survey simulator is run until the number of detections matches the real detections (1 or 3, depending on the $\Delta v$  range).
The survey simulator was run 2,000 separate times for each input size distribution.
This provides a measurement of the range of intrinsic population estimates consistent with the real detections.
The median of these individual estimates is the reported population estimate, and the 2$\sigma$ uncertainties were determined based on the value of the outlying 2.5\% of estimates (high and low).
The input distribution has $H_r<9.5$, as the OSOSS Ensemble was sensitive to this magnitude limit.
We also report a population estimate for $H_r<6$ ($\sim91$~km, assuming albedo 0.85), which was calculated from the $H_r<9.5$ ($\sim18$~km) population estimate and the known size distribution slope for each model.
These population estimates are reported in Table \ref{popEst}.
We also require the OSSOS population estimate should be consistent with at least the number of identified Haumea family members, 7 objects  $\Delta v$ $<$100 m s$^{-1}$ and 22 objects with  $\Delta v$ $<$160 m s$^{-1}$, as a constraint on the acceptable size distributions.

\newpage

\noindent
\\ {\bf Correspondence.} Correspondence should be addressed to Rosemary E. Pike. 
\\ {\bf Acknowledgements.} Based on observations obtained with MegaPrime/MegaCam, a joint project of CFHT and CEA/DAPNIA, at the Canada-France-Hawaii Telescope (CFHT) which is operated by the National Research Council (NRC) of Canada, the Institut National des Sciences de l'Univers of the Centre National de la Recherche Scientifique of France, and the University of Hawaii. 
The authors wish to recognize and acknowledge the very significant cultural role of the summit of Maunakea.  We are most fortunate to have the opportunity to conduct observations from this mountain.
BCNP, DR, and SM acknowledge support from a BYU Mentored Environment Grant.
\\ {\bf Author Contributions.} REP tested the models using the survey simulator, determined the mass and population estimate, and wrote the majority of the paper draft.  BCNP generated the orbital distribution models used, and SM and DR and the classification of objects as Haumea family member candidates.
MA assisted with the mass estimate and generating the approximate Pan-STARRS1 survey simulator blocks.
MA, MB, YTC, BJG, JJK, SG and KV did the object detections and survey characterization for the OSSOS survey.
\\ {\bf Competing Interests.} The authors declare that they have no competing financial interests.
\\ {\bf Data Availability.} The data that support the plots within this paper are available from the corresponding author upon reasonable request.
\\ {\bf Code Availability.} The Survey Simulator is available publicly from the OSSOS webpages: \textit{http://www.ossos-survey.org/simulator.html}.  A detailed description of usage is available \citep{lawlerFASS}.

\clearpage


\end{document}